\documentclass[12pt]{iopart}

\usepackage{iopams}

\usepackage[colorlinks=true, pdfstartview=FitV, linkcolor=blue, citecolor=black, urlcolor=black, breaklinks]{hyperref}

\usepackage{mathabx} 
\usepackage{color}
\usepackage{graphicx}

\begin{document}

\title[Earth satellite dynamics by Picard iterations]{Earth satellite dynamics by Picard iterations}

\author{M Lara}

\address{SCoTIC--University of La Rioja, Madre de Dios, 53, 26006 Logro\~no, La Rioja, Spain}
\ead{\mailto{martin.lara@unirioja.es}, \mailto{mlara0@gmail.com}}
\vspace{10pt}
\begin{indented}
\item[] \color{red}draft of April 14, 2022 
\end{indented}

\begin{abstract}
The main effects of the Earth's oblateness on the motion of artificial satellites are usually derived from the variation of parameters equations of an average representation of the oblateness disturbing function. Rather, we approach their solution under the strict mathematical assumptions of Picard's iterative method. Our approach recovers the known linear trends of the right ascension of the ascending node and the argument of the perigee, but differs from the accepted solution in the value of the mean motion. This amended rate radically improves the in-track errors of typical orbit propagations. In addition, our truncation of the Picard iterations solution to its secular terms includes the corrections that must be applied to the osculating initial conditions in the right propagation of the mean dynamics. 
\end{abstract}

%
\vspace{2pc}

%
%
%
%

\section{Introduction}

Most introductory textbooks on celestial mechanics and astrodynamics present the derivation of Lagrange's planetary equations in the context of general perturbations and the variations of the parameters method (see \cite[\S11]{Danby1992}, \cite[\S7]{Roy2005}, for instance). Their solution is approached by successive approximations, but going beyond the first iteration is usually declined due to the complexities of the formulation. Moreover, when dealing with the dynamics of artificial satellites of the Earth, the main effects are customarily derived from an \emph{averaged} form of the variation of parameters equations \cite[\S10.6]{Battin1999} \cite[\S9.1]{Vallado2013}. However, we will see that an improved description of these effects is obtained with little effort from the straightforward solution of the non-averaged equations based on Picard's method.
\par

Artificial satellite theory is characterized by the dominant effect of the zonal harmonic of the second degree on the dynamics of low Earth orbits. This fact brought the so-called main problem of artificial satellite theory, or $J_2$--problem, to analogous levels of popularity to those of the restricted three-body problem at the beginning of the space era. The $J_2$--problem admits the energy and the third component of the angular momentum vector as independent integrals, but the third integral that would guarantee the existence of a closed form solution does not exist beyond particular integrable cases \cite{Garfinkel1958,Vinti1961,Jezewski1983}. Numerical explorations using unrealistic high values of $J_2$ unveiled chaotic regions for certain values of the energy, thus pointing to the non-integrable character of the $J_2$-problem \cite{Danby1968,Broucke1994}, a conjecture that was soon supported analytically \cite{IrigoyenSimo1993,CellettiNegrini1995}. However, for the small value of the Earth's $J_2$ coefficient, the size of the regions in which chaos may emerge is so small that the lack of integrability can be completely ignored in practice \cite{Simo1991}, and existing analytical perturbation solutions of the main problem may remain accurate to machine precision over long time intervals \cite{Lara2020}.
\par

On the other hand, the limitations inherent to the $J_2$-problem model in the description of the dynamics of Earth's artificial satellites  make that highly accurate solutions are of limited interest in practice due to the length of the series involved in the representation of the solution \cite{CoffeyDeprit1982,Healy2000}. Therefore, they are replaced by much simpler analytical solutions that just capture the bulk of the main problem dynamics. These kinds of solutions are generally encompassed under the name of intermediary orbits (see \cite[\S11.5]{Roy2005}, \cite{Deprit1981} and references therein). More precisely, good main problem intermediaries must yield, at least, the same average dynamics as the main problem up to first order effects of $J_2$. That is, on average, the solution must bring the orbital plane to bear a small linear variation of the right ascension of the ascending node with constant inclination, and force a small but steady motion of the argument of the perigee in the orbital plane. The former vanishes for polar orbits, whereas the later does it at the critical inclinations of $90\pm26.6$ degrees. The rates of variation of these angles are $\mathcal{O}(J_2)$ when compared with the orbital mean motion, which, on average, differs from the Keplerian rate on a linear term which is also $\mathcal{O}(J_2)$.
\par

As mentioned before, these general characteristics resulting from the oblateness perturbation are usually derived from an average representation of the disturbing function \cite[\S10.6]{Battin1999}, \cite[\S5.3.1]{Lara2021}, and can be used as a seed for integrating more accurate solutions that include periodic effects \cite[\S3.4]{Kaula1966}. However, in spite of the success of this procedure in constructing accurate transformations between osculating and mean elements \cite{Guinn1991}, it fails in predicting the time history of the mean anomaly within a reasonable accuracy save for particular configurations. On the contrary, Picard's constructive proof for the existence and unicity of solutions to ordinary differential equations (see \cite[\S7]{Hurewicz1958} or \cite[\S1.6]{PuigAdam1967}, for instance) furnish us with a rigorous mathematical procedure for approaching the closed form solution of the $J_2$-problem by iterations. Due to the different time scales in which the angles evolve in a perturbation problem, the variation equations are preferably formulated using strict elements as well as a fictitious time like the independent variable \cite[\S11.4.1]{Roy2005}. Then, once the differential system is solved in closed form of the eccentricity up to some iteration of Picard's method, the mean anomaly is in turn computed by indefinite integration.
\par

Picard's iterative method is commonly objected in the solution of Lagrange's planetary equations due to the complexities in progressing beyond the first iteration, as well as the appearance of mixed secular periodic terms as byproduct of the required expansions, cf.~\cite[\S7.7.2]{Roy2005} or \cite[\S6.7]{BoccalettiPucacco1998v2}.  Certainly, the trouble happens in the solution of the main problem with Picard's method too, thus constraining in practice the accuracy of the analytical solution to the first order effects of the oblateness coefficient. Still, we can avoid the appearance of mixed terms in a second iteration proceeding with some care, in the fashion of Kaula's linear theory \cite[\S3.4]{Kaula1966}. In this way the Picard iterations solution of the $J_2$-problem can cope with non-resonant long-period terms, which allow it to remain within an acceptable accuracy for longer time intervals.
\par

Therefore, we find Picard's method useful in the solution of the main problem of artificial satellite theory in three different facets: i) the method is very easy to grasp with the standard undergraduate background on differential equations; ii) it leads naturally to the formulation of the variation equations in a fictitious time, in this way illustrating at a very elemental level the benefits of regularization techniques, which apply both to numerical and analytical integration schemes \cite{StiefelScheifele1971,ScheifeleGraf1974,PelaezHedoRodriguez2007,Herreraetal2014,Roa2017,Lara2017if,Lara2022}; iii) the solution can be arranged in the form of secular and periodic terms, thus building a bridge with the basic ideas of perturbation theory.
\par

In particular, our solution recovers the known linear trends in the long-term motion of the node and perigee, and provides a simple way of illustrating that the mean dynamics arising from a given initial state must be propagated from a different set of initial conditions than the osculating ones ---a fact that seems to be well understood only within the scope of perturbation solutions (see discussions in \cite[\S9.9.1]{Vallado2013}).
Moreover, our Picard iterations solution discloses an additional term to the secular variation of the mean anomaly commonly reported in the literature. Most notably, this new term prevents the abnormal growth of errors in the in-track direction, in this way avoiding the need of calibration of the mean motion that is typical of perturbation solutions \cite{LyddaneCohen1962,BreakwellVagners1970,Lara2020arxiv}. Because of that, and due to the fact that it is made only of elementary functions, the new analytical solution provides an appealing alternative of comparable accuracy to usual main problem intermediaries in the literature.
\par

It is well known that the standard set of Keplerian orbital elements is singular for specific orbit configurations, and in particular for circular orbits. In spite of that, we adhere to the tradition and integrate the differential equations of the flow in these elements for their immediate insight, but also for easing comparison with alternative solutions in the literature. Offending divisors for the lower eccentricity orbits can always be avoided by analogously approaching by Picard iterations the solution of the variation equations of a different set of variables that may avoid the undesired singularities \cite{Hintz2008}.
\par

\section{The main problem Hamiltonian and the variation equations}

The main problem Hamiltonian is obtained by making zero all the harmonic coefficients of the usual expansion of the geopotential except for the zonal harmonic of the second degree. When this truncation is written in Cartesian variables, we obtain
\begin{equation} \label{HamCart}
\mathcal{H}=\frac{1}{2}\left(X^2+Y^2+Z^2\right)-\frac{\mu}{r}
-J_2\frac{\mu}{r}\frac{R_\Earth^2}{r^2}\left(\frac{1}{2}-\frac{3}{2}\frac{z^2}{r^2}\right),
\end{equation}
where the conjugate momenta $(X,Y,Z)$ to the Cartesian variables $(x,y,z)$ coincide with the velocity, $\mu$ is the Earth's gravitational parameter, $R_\Earth$ is the Earth's equatorial radius, $J_2=-C_{2,0}\approx10^{-3}$ is the Earth's zonal harmonic coefficient of the second degree, and $r=(x^2+y^2+z^2)^{1/2}$.
\par

The differential equations of the flow stemming from (\ref{HamCart}), that is, the Hamilton equations, provide a simple and efficient formulation for the numerical integration of the main problem dynamics. However, the formulation in coordinates is not suitable in general for the search of analytical solutions, in which case the formulation in elements is preferred. When the usual Keplerian set $(a,e,I,\Omega,\omega,M)$ is used, Hamiltonian (\ref{HamCart}) takes the compact, meaningful form (see \cite{Kozai1959}, for instance, where $A_2=\frac{3}{2}J_2R_\Earth^2$)
\begin{equation} \label{Hamorb}
\mathcal{H}=-\frac{\mu}{2a}
-\frac{1}{4}J_2\frac{\mu}{r}\frac{R_\Earth^2}{r^2}\left[2-3s^2+3s^2\cos(2f+2\omega)\right],
\end{equation}
in which we abbreviated $s\equiv\sin{I}$, the radius $r$ is replaced by the conic equation $r=p/(1+e\cos{f})$, with the orbit parameter $p=a(1-e^2)$, and, due to the Hamiltonian formulation, all symbols must be considered \emph{functions} of some set of canonical variables. In particular, when using the Delaunay canonical variables $(\ell,g,h,L,G,H)$ we have $a=a(L)\equiv{L}^2/\mu$, $p=p(G)\equiv{G}^2/\mu$, $\omega=\omega(g)\equiv{g}$, $e=e(G,L)\equiv(1-G^2/L^2)^{1/2}$, $s=s(G,H)\equiv(1-H^2/G^2)^{1/2}$, and $f={f}(\ell,L,G)$ is an implicit function of the mean anomaly $\ell=\ell(M)\equiv{M}$ that involves the solution of the Kepler equation \cite{Delaunay1860}. We note that the cyclic character of the right ascension of the ascending node $\Omega=\Omega(h)\equiv{h}$, as results from its absence of the Hamiltonian, turns the polar component of the angular momentum $H$ into an integral of the main problem dynamics.
\par

The time variation of the Delaunay variables is obtained from corresponding Hamilton equations. Namely,
\begin{equation} \label{HamiltonEqs}
\frac{\rmd (\ell,g,h)}{\rmd t}=\frac{\partial\mathcal{H}}{\partial(L,G,H)}, \qquad
\frac{\rmd (L,G,H)}{\rmd t}=-\frac{\partial\mathcal{H}}{\partial(\ell,g,h)}.
\end{equation}
Rather, for better insight and comparison with usual results in the literature, we use the chain rule to compute the variations of the orbital elements. Hence,
\begin{equation} \label{poisson}
\frac{\rmd \xi}{\rmd t}=
\frac{\partial\xi}{\partial\ell}\frac{\partial\mathcal{H}}{\partial{L}}
+\frac{\partial\xi}{\partial{g}}\frac{\partial\mathcal{H}}{\partial{G}}
+\frac{\partial\xi}{\partial{h}}\frac{\partial\mathcal{H}}{\partial{H}}
-\frac{\partial\xi}{\partial{L}}\frac{\partial\mathcal{H}}{\partial\ell}
-\frac{\partial\xi}{\partial{G}}\frac{\partial\mathcal{H}}{\partial{g}}
-\frac{\partial\xi}{\partial{H}}\frac{\partial\mathcal{H}}{\partial{h}}
= \{\xi;\mathcal{H}\},
\end{equation}
where $\xi$ is an arbitrary function of the Delaunay variables, and the curly brackets represent the Poisson bracket. We obtain the usual result, which we rather express in the form
%
\begin{eqnarray}
\label{ap}
\frac{\rmd a}{\rmd t} &=& \frac{J_2}{4}\frac{G}{r^2}\frac{R_\Earth^2}{p^2}\frac{a}{\eta^2}\left[
\frac{3s^2-2}{2}\sum_{j=1}^3a_{0,j}\sin{j}f-\frac{s^2}{4}\sum_{j=-1}^5a_{1,j}\sin(jf+2\omega) \right], \\ \label{ep}
\frac{\rmd e}{\rmd t} &=& \frac{J_2}{4}\frac{G}{r^2}\frac{R_\Earth^2}{p^2}\left[
\frac{3s^2-2}{4}\sum_{j=1}^3e_{0,j}\sin{j}f-\frac{s^2}{8}\sum_{j=-1}^5e_{1,j}\sin(jf+2\omega) \right], \\ \label{Ip}
\frac{\rmd I}{\rmd t} &=&
-\frac{J_2}{4}\frac{G}{r^2}\frac{R_\Earth^2}{p^2}cs\sum_{j=1}^3I_{1,j}\sin(jf+2\omega), \\ \label{hp}
\frac{\rmd \Omega}{\rmd t} &=& \frac{J_2}{4}\frac{G}{r^2}\frac{R_\Earth^2}{p^2}c\left[
-6(1+e\cos{f})+\sum_{j=1}^3\Omega_{1,j}\cos(jf+2\omega)\right], \\ \nonumber
\frac{\rmd \omega}{\rmd t} &=& \frac{J_2}{4}\frac{G}{r^2}\frac{R_\Earth^2}{p^2}\left[-3(5s^2-4)(1+e\cos{f})
-\frac{3s^2-2}{4e}\sum_{j=1}^3\omega_{0,j}\cos{j}f \right. \\ \label{gp} && \left.
+\frac{1}{8e}\sum_{j=-1}^5\sum_{k=0}^{j\bmod2}\omega^*_{j,k}e^{1+(j\bmod2)-2k}\cos(jf+2\omega)\right], \\ \nonumber
\frac{\rmd M}{\rmd t} &=& \; n +\frac{J_2}{4}\frac{G}{r^2}\frac{R_\Earth^2}{p^2}\eta\left[-3(3s^2-2)(1+e\cos{f})
+\frac{3s^2-2}{4e}\sum_{j=1}^3\omega_{0,j}\cos{j}f \right. \\ \label{Mp} && \left.
-\frac{1}{8e}\sum_{j=-1}^5\sum_{k=0}^{j\bmod2}M^*_{j,k}e^{1+(j\bmod2)-2k}\cos(jf+2\omega)\right],
\end{eqnarray}
in which $\eta=\left(1-e^2\right)\!{}^{1/2}$, $c=\left(1-s^2\right)\!{}^{1/2}$, $n=\left(\mu/a^3\right)\!{}^{1/2}$, the eccentricity polynomials $q_{i,j}$, $q\in(a,e,I,\Omega,\omega)$ are given in Table~\ref{t:eccpo}, the inclination polynomials $Q_{i,j}$, $Q\in(\omega^*,M^*)$ are given in Table~\ref{t:incpo}, and, in preparation of following developments, we keep the factor $G/r^2$ in each variation equation. Recall that the Delaunay variable $G$ is the specific angular momentum. 
\par

\begin{table}[htb]
\caption{\label{t:eccpo}Eccentricity polynomials $q_{i,j}$ in (\protect\ref{ap})--(\protect\ref{Mp}). }
\footnotesize
\begin{tabular}{@{}lccccccccc@{}} 
\br
$q\,\backslash\;{}_{i,j}$ & ${}_{0,1}$ & ${}_{0,2}$ & ${}_{0,3}$ 
 & ${}_{1,-1}$ & ${}_{1,1}$ & ${}_{1,2}$ & ${}_{1,3}$ & ${}_{1,4}$ & ${}_{1,5}$ \\
\mr 
$a$ & $12e + 3e^3$ & $12e^2$ & $3e^3$
    & $-3 e^3$ & $9 e^3+36 e$ & $72e^2+48$ & $27e^3+108e$ & $72 e^2$ & $15e^3$ \\
$e$ & $12+3e^2$ & $12e$ & $3e^2$
    & $-3e^2$ & $33e^2+12$ & $120e$ & $51e^2+84$ & $72e$ & $15e^2$ \\
$I$ & $-$ & $-$ & $-$ & $-$ & $3e$ & $6$ & $3e$ & $-$ & $-$ \\
$\omega$ & $12-3e^2$ & $12e$ & $3e^2$ & $-$ & $-$ & $-$ & $-$ & $-$ & $-$ \\
\br 
\end{tabular}\\
$q_{1,0}=0$ and $\Omega_{i,j}=I_{i,j}$.
\end{table}
\normalsize
\begin{table}[htb]
\caption{\label{t:incpo}Inclination polynomials $Q_{i,j}$ in (\protect\ref{gp}) and (\protect\ref{Mp}). }
\footnotesize
\begin{tabular}{@{}lccccccccc@{}} 
\br
$Q\,\backslash\;{}_{i,j}$  & ${}_{-1,0}$ & ${}_{1,0}$ & ${}_{1,1}$ 
 & ${}_{2,0}$ & ${}_{3,0}$ & ${}_{3,1}$ & ${}_{4,0}$ & ${}_{5,0}$ \\
\mr 
$\;\omega^*$ & $3s^2$ & $45s^2-24$ & $-12s^2$ & $120s^2-48$ & $57s^2-24$ & $84s^2$ & $72s^2$ & $15s^2$ \\
$M^*$    & $3s^2$ & $-51s^2$   & $-12s^2$ & $-72s^2$    & $-39s^2$   & $84s^2$ & $72s^2$ & $15s^2$ \\
\br 
\end{tabular}
\end{table}
\normalsize

The variations of the argument of the perigee and the mean anomaly are singular for circular orbits due to the appearance of the eccentricity as a divisor. But this is just a consequence of using Keplerian elements that is easily avoided by computing the variations of a different set of non-singular variables \cite{Lyddane1963,DepritRom1970,ArsenaultFordKoskela1970}. For instance, the variation of the mean distance to the node $F=M+\omega$ is obtained by addition of (\ref{gp}) and (\ref{Mp}). Namely,
\begin{eqnarray} \nonumber
\frac{\rmd F}{\rmd t}&=& \; n
-\frac{J_2}{4}\frac{G}{r^2}\frac{R_\Earth^2}{p^2}\Bigg\{3\left[(3s^2-2)\eta+5s^2-4\right](1+e\cos{f})
+\frac{3s^2-2}{4(1+\eta)} \\ \label{dFdt}  &&
\times{e}\sum_{j=1}^3\omega_{0,j}\cos{j}f
+\frac{1}{8}\sum_{j=-1}^5\sum_{k=0}^{j\bmod2}P_{j,k}(e,s)\cos(jf+2\omega)\Bigg\},
\end{eqnarray}
where it is easy to check that the coefficients $P_{j,k}\equiv(\eta{M}^*_{j,k}-\omega^*_{j,k})e^{(j\bmod2)-2k}$ are free from division by the eccentricity.
\par

On the other hand, the singularity for circular orbits is not of concern for illustrating our findings, and hence we adhere to the tradition of using Keplerian elements for better insight as well as to ease comparison with alternative solutions in the literature.
\par

\section{Picard's iterative approach}

The computation of an analytical solution to (\ref{ap})--(\ref{Mp}) is naturally approached with Picard's method, which we outline in what follows for completeness. Let
\begin{equation} \label{fods}
\frac{\rmd \xi_i}{\rmd \tau}=\chi_i(\xi_j,\tau), \quad \xi_i(\tau_0)=\xi_{i,0}, \qquad i,j=1,\dots m,
\end{equation}
be a first order differential system in which $\tau$ is the independent variable and $\xi_i$ are $m$ dependent variables. Assuming that the conditions that guarantee the existence and uniqueness of the solution apply, we want to compute a particular solution for the initial conditions $\xi_{i,0}$. If that solution, say $\xi_i=\xi_i(\tau,\xi_{j,0})$, were already known, by replacing it into (\ref{fods}) we obtain the identity
\[
\frac{\rmd \xi_i}{\rmd \tau}=\chi_i(\xi_j(\tau,\xi_{j,0}),\tau),  \qquad i,j=1,\dots m,
\]
which can be rewritten in the integral form
\begin{equation} \label{identity}
\xi_i=\xi_{i,0}+\int_{\tau_0}^\tau\chi_i(\xi_j(\tau,\xi_{j,0}),\tau)\,\rmd \tau, \qquad i,j=1,\dots m.
\end{equation}
\par

When the solution to (\ref{fods}) is unknown (\ref{identity}) still applies, but now it becomes a formal quadrature rather than an identity. Then, assuming the analytic character of the $\chi_i\equiv\chi_i(\xi_{j},\tau)$, we can replace them by their Taylor series expansions
\begin{eqnarray} \nonumber
\chi_i &=& \chi_i(\xi_{j,0},\tau_0)+\left(\frac{\partial\chi_i}{\partial\tau}+\sum_{j=1}^m\frac{\partial\chi_i}{\partial\xi_j}\frac{\rmd \xi_j}{\rmd \tau}\right)\bigg|_{\xi_{j,0},\tau_0}\Delta\tau + \mathcal{O}(\Delta\tau)^2 \\ \label{chiseries}
&=& \chi_i(\xi_{j,0},\tau_0)+\left.\frac{\partial\chi_i}{\partial\tau}\right|_{\tau_0}\!\Delta\tau+\sum_{j=1}^m\left.\frac{\partial\chi_i}{\partial\xi_j}\right|_{\xi_{j,0},\tau_0}\left(\left.\frac{\rmd \xi_j}{\rmd \tau}\right|_{\tau_0}\!\Delta\tau\right) + \mathcal{O}(\Delta\tau)^2.
\end{eqnarray}
That is,
\begin{equation} \label{chip}
\chi_i=\chi_i(\xi_{j,0},\tau)+\sum_{j=1}^m\left.\frac{\partial\chi_i}{\partial\xi_j}\right|_{\xi_{j,0},\tau_0}(\xi_j-\xi_{j,0}) + \mathcal{O}(\Delta\tau)^2,
\end{equation}
which shows the connection of the procedure with the method of successive approximations \cite[\S11.11]{Danby1992}.
\par

In (\ref{chip}) the dependence of $\xi_j=\xi_j(\tau,\xi_{j,0})$ on the independent variable and the initial conditions is not known hitherto. However, if we constrain to such interval $\Delta\tau=\tau-\tau_0$ that the differences $\xi_j-\xi_{j,0}$ remain small enough, we can neglect these differences and compute a first analytical approximation to the solution $\xi_i\approx\xi_{i,1}(\tau;\xi_{j,0})$ by solving
\begin{equation} \label{pifiit}
\xi_{i,1}=\xi_{i,0}+\int_{\tau_0}^\tau\chi_i(\xi_{j,0},\tau)\,\rmd \tau.
\end{equation}
\par

A refinement of this solution is sometimes computed by replacing $\xi_j$ by $\xi_{j,1}$ into (\ref{chip}). That is,
\[
\chi_i\approx\chi_i(\xi_{j,0},\tau)+\sum_{j=1}^m\left.\frac{\partial\chi_i}{\partial\xi_j}\right|_{\xi_{j,0}}\int_{\tau_0}^\tau\chi_j(\xi_{k,0},\tau)\,\rmd \tau,
\]
which is in turn replaced into (\ref{identity}) and solved by indefinite integration, cf.~\cite{Kozai1959,Kaula1966}. Rather, the analytical solution is constructed stepwise with Picard's iterative method. That is, the solution to (\ref{fods}) is obtained from the sequence
\begin{equation} \label{newrap}
\xi_{i,k}=\xi_{i,0}+\int_{\tau_0}^\tau\chi_i[\xi_{j,k-1}(\tau,\xi_{j,0}),\tau]\,\rmd \tau,
\end{equation}
which starts from (\ref{pifiit}). The demonstration of the convergence of Picard's method under some general conditions can be found, for instance, in \cite{Hurewicz1958,PuigAdam1967}.
\par

\section{Variation equations in the fictitious time}

Picard's iterative scheme seems appealing for computing an analytical solution of (\ref{ap})--(\ref{Mp}) based on the constants of the Keplerian solution. Indeed, in (\ref{chiseries}) we check that, for the strict elements, the first two summands ---or the first term of (\ref{chip})--- are $\mathcal{O}(J_2)$, whereas the following summation would contribute terms $\mathcal{O}(J_2^2)$, and the neglected terms would be of higher order.
\par

Then, on account of $G/r^2=n\eta(a/r)^2$, the domain in which the Picard iterations solution applies is, in fact, bounded by the condition $J_2n\Delta{t}<1$. However, this is not the case of the variation of the mean anomaly, which grows fast as compared with the variations of the strict elements, and hence the assumption $M-M_0$ small will constrain the validity of the solution to much shorter times. Therefore, when using this computational scheme it is better to replace $M$ by a slowly evolving variable. In view of the form of (\ref{Mp}), a natural choice is the time element $\beta$ defined by the differential relation, cf.~\cite[Eq.~(10.39)]{Battin1999},
\begin{equation} \label{sigma}
\frac{\rmd \beta}{\rmd t}=\frac{\rmd M}{\rmd t}-\sqrt{\frac{\mu}{a^3}}.
\end{equation}
With this choice, we replace (\ref{Mp}) by
\begin{eqnarray} \nonumber
\frac{\rmd \beta}{\rmd t}&=& 
\frac{J_2}{4}\frac{G}{r^2}\frac{R_\Earth^2}{p^2}\left[-3(3s^2-2)(1+e\cos{f})
+\frac{3s^2-2}{4e}\eta\sum_{j=1}^3\omega_{0,j}\cos{j}f \right. \\ \label{bp} && \left.
-\frac{\eta}{8e}\sum_{j=-1}^5\sum_{k=0}^{j\bmod2}M^*_{j,k}e^{1+(j\bmod2)-2k}\cos(jf+2\omega)\right].
\end{eqnarray}
\par

Still, the mean anomaly is present in the variation equations through its implicit dependence  on $f$. Then, in order for the variation equations to take a form amenable to solution by Picard iterations, we carry out the integration in a fictitious time $\tau$ given by the differential relation
\begin{equation} \label{dtf}
\rmd t=\frac{r^2}{G}\,\rmd \tau,
\end{equation}
which makes now evident why we left the factor $G/r^2$ explicit in (\ref{ap})--(\ref{Mp}) and (\ref{bp}). When doing so, the law of areas shows that we are assigning to the argument of latitude the role of the fictitious time. On the other hand, since the variation equations are already $\mathcal{O}(J_2)$, if we constrain the solution to the first order of $J_2$ effects, then the fictitious time evolves at the same rate as the true anomaly, which, therefore, can be taken as the independent variable \cite[\S11.4.1]{Roy2005}. Then, the variations of the orbital elements in the fictitious time become
\begin{equation} \label{dxdf}
\frac{\rmd(a,e,I,\Omega,\omega,\beta)}{\rmd f}=\frac{r^2}{G}\frac{\rmd(a,e,I,\Omega,\omega,\beta)}{\rmd t}, \end{equation}
where the time derivatives of the elements $(a,e,I,\Omega,\omega,\beta)$ are replaced by (\ref{ap})--(\ref{gp}) and (\ref{bp}), respectively. Once (\ref{dxdf}) is solved in the fictitious time up to some iteration $k$ of Picard's method, the time history of the mean anomaly is obtained from (\ref{sigma}) as
\begin{equation} \label{Moft}
M=M_0+\beta_k(t)-\beta_k(t_0)+\sqrt{\mu}\int_{t_0}^t\frac{1}{a_k(t)^{3/2}}\,\rmd t.
\end{equation}
\par

The relation between the physical and fictitious times is obtained from the integration of (\ref{dtf}). Namely,
\begin{equation} \label{foft}
t=t_0+\int_{f_0}^f\frac{\left[\left(1-e_k^2\right)a_k\right]^{3/2}}{(1+e_k\cos{f})^2\mu^{1/2}}\,\rmd f.
\end{equation}
As expected, when $k=0$ we recover the Kepler equation $t=t^*+(u-e\sin{u})/n$, where $u=u(f,e)$ is the eccentric anomaly and $t^*=t_0-(u_0-e\sin{u}_0)/n$ is the time of perigee passage.
\par

\section{First Picard iteration}

Assume that the elements in the right side of the differential system (\ref{dxdf}) remain constant as given by the initial conditions $a_0$, $e_0$, $I_0$, $\Omega_0$, $\omega_0$, and $\beta_0=0$. In addition, $M_0=M(t_0)$ and $f_0=f(M_0,e_0)$. Then, the first iteration of Picard's method is readily obtained from (\ref{pifiit}). For the sake of easing comparisons with existing expressions in the literature, we split the solution into secular and purely short-period terms in the \emph{mean} anomaly, and write it in the form
\begin{eqnarray} \label{a1}
a_1 &=& a_0 +a_0\varepsilon\left[a_{1,\mathrm{P}}(f)-a_{1,\mathrm{P}}(f_0)\right],
\\ \label{e1}
e_1 &=& e_0 +\varepsilon\left[e_{1,\mathrm{P}}(f)-e_{1,\mathrm{P}}(f_0)\right],
\\ \label{I1}
I_1 &=& I_0 +\varepsilon{c}\left[I_{1,\mathrm{P}}(f)-I_{1,\mathrm{P}}(f_0)\right],
\\ \label{h1} 
\Omega_1 &=& \Omega_0 -6\varepsilon{c}(M-M_0) +\varepsilon{c}\left[\Omega_{1,\mathrm{P}}(f)-\Omega_{1,\mathrm{P}}(f_0)\right],
\\ \label{g1}
\omega_1 &=& \omega_0 -3\varepsilon(5s^2-4)(M-M_0) +\varepsilon\left[\omega_{1,\mathrm{P}}(f)-\omega_{1,\mathrm{P}}(f_0)\right],
\\ \label{s1}
\beta_1 &=& -3\varepsilon\eta(3s^2-2)(M-M_0) +\varepsilon\left[\beta_{1,\mathrm{P}}(f)-\beta_{1,\mathrm{P}}(f_0)\right],
\end{eqnarray}
where we abbreviate
\begin{equation}
\varepsilon=\frac{1}{4}J_2\frac{R_\Earth^2}{p^2},
\end{equation}
and,
\begin{eqnarray} \label{a1s}
a_{1,\mathrm{P}} &=& 
-\frac{3s^2-2}{2\eta^2}\sum_{j=0}^3\tilde{a}_{0,j}\cos{j}f
+\frac{s^2}{4\eta^2}\sum_{j=-1}^5\tilde{a}_{1,j}\cos(jf+2\omega), \\ \label{e1s}
e_{1,\mathrm{P}} &=& 
-\frac{3s^2-2}{4}\sum_{j=0}^3\tilde{e}_{0,j}\cos{j}f+\frac{s^2}{8}\sum_{j=-1}^5\tilde{e}_{1,j}\cos(jf+2\omega), \\ \label{i1s}
I_{1,\mathrm{P}} &=& s\sum_{j=0}^3\tilde{I}_{1,j}\cos(jf+2\omega), \\ \label{h1s}
\Omega_{1,\mathrm{P}} &=& 
-6(f-M+e\sin{f})+\sum_{j=0}^3\tilde\Omega_{1,j}\sin(jf+2\omega), \\ \nonumber
\omega_{1,\mathrm{P}} &=& -3(5s^2-4)(f-M+e\sin{f})
-\frac{3s^2-2}{4e}\sum_{j=1}^3\tilde\omega_{0,j}\sin{j}f \\ \label{g1s} &&
+\frac{1}{8e}\sum_{j=-1}^5\sum_{k=0}^{j\bmod2}\tilde\omega^*_{j,k}e^{1+(j\bmod2)-2k}\sin(jf+2\omega), \\ \nonumber
\beta_{1,\mathrm{P}} &=& -3\eta(3s^2-2)(f-M+e\sin{f})
+\frac{3s^2-2}{4e}\eta\sum_{j=1}^3\tilde{\omega}_{0,j}\sin{j}f \\ \label{l1s} &&
-\frac{\eta}{8e}\sum_{j=-1}^5\sum_{k=0}^{j\bmod2}\tilde{M}^*_{j,k}e^{1+(j\bmod2)-2k}\sin(jf+2\omega).
\end{eqnarray}
in which $\tilde{q}_{i,j}=q_{i,j}/j$ if $j\ne0$, and the coefficients $\tilde{q}_{i,0}$, which are listed in Table~\ref{t:adincpo}, appear as a consequence of the introduction in the integration of such arbitrary constants that cancel long-period terms stemming from the closed form integration in the fictitious time.
\par
\begin{table}
\caption{\label{t:adincpo}Additional inclination functions in (\protect\ref{a1s})--(\protect\ref{l1s}).}
\footnotesize
\begin{tabular}{@{}lccccccccc@{}} 
\br 
$\tilde{a}_{1,0}=$&$18e^2$ & $\tilde{a}_{0,0}=$&$-4\eta^3-6\eta^2+10$ \\
$\tilde{e}_{1,0}=$&$-2e(8\eta^3-5\eta^2-18\eta-9)/(1+\eta)^2$ & $\tilde{e}_{0,0}=$&$10e+4e\eta^2/(1+\eta)$ \\
$\tilde{I}_{1,0}=$&$e^2(1+2\eta)/(1+\eta)^2$  \\
$\tilde\omega^*_{1,0}=$&$2s^2-8-8\eta^2\left[\eta(4s^2-2)+3 s^2-2\right]/(1+\eta)^2$  \\
$\tilde{M}^*_{1,0}=$&$2s^2(16\eta^3+5\eta^2-30\eta-15)/(1+\eta)^2$ \\
\br 
\end{tabular}\\
$\tilde\Omega_{1,0}=\tilde{I}_{1,0}$
\end{table}
\normalsize

Taking into account that the equation of the center $f-M$ is a purely periodic function of $M$ \cite[\S{II.6}]{BrouwerClemence1961}, and using the known relations \cite{Kozai1962AJ,Kelly1989}
\[
\frac{1}{2\pi}\int_0^{2\pi} {{\sin(mf+\alpha)}\choose{\cos(mf+\alpha)}}\,\rmd M=
\frac{(-e)^m}{(1+\eta)^m}(1+m\eta) {{\sin\alpha}\choose{\cos\alpha}},
\]
we readily check that in the time in which the mean anomaly advances by $2\pi$ (\ref{a1s})--(\ref{l1s}) average out to zero, and hence their purely periodic character. There is no surprise in checking that the short-period corrections are the same as those in \cite[Eq.~(10)]{Kozai1959} after removing there the hidden long-period terms that are due to the closed form formulation in \cite[Eq.~(12)]{Kozai1959} (see also \cite[\S11.4.1]{Roy2005}).
\par

The true anomaly corresponding to a given physical time is obtained in implicit form after integrating (\ref{foft}). The closed form integration would be possible after expanding the integrand in powers of $J_2$ and truncating the expansion to the first order. However, this procedure is not needed because, as follows from (\ref{a1})--(\ref{I1}), $f$ only enters the solution through terms that already are $\mathcal{O}(J_2)$, and, therefore, it can be solved directly from the Kepler equation, in which we make $M=\left(\mu/a_0^3\right)\!{}^{1/2}(t-t^*)$.
\par

\subsection{The mean anomaly at the first iteration}

To complete the solution it only remains to solve the indefinite integral in (\ref{Moft}) replacing $a_k=a_1$ by (\ref{a1}). Since the closed form solution of this integral is not known, we rather expand the integrand in powers of $J_2$, which are truncated to the first order in agreement with our previous assumptions. Namely,
\[
\frac{1}{a_1^{3/2}}=\frac{1}{a_0^{3/2}}\left\{1-\frac{3}{2}\varepsilon\left[a_{1,\mathrm{P}}(f)-a_{1,\mathrm{P}}(f_0)\right]+\mathcal{O}\left(J_2^2\right)\right\}.
\]
\par

Then, due to the form of $a_{1,\mathrm{P}}$ in (\ref{a1s}), which only involves cosine functions of $f$ and the constant initial values of the osculating elements, the closed form integration of (\ref{Moft}) is readily achieved. Indeed, on the one hand, terms independent of the true anomaly are trivially integrated by just multiplying them by the physical time $t$. Thus,
\[
\mathcal{I}\equiv\int_{t_0}^t\sqrt{\frac{\mu}{a_1^3}}\,\rmd t=
n\left[1+\frac{3}{2}\varepsilon a_{1,\mathrm{P}}(f_0)\right](t-t_0)-
\frac{3}{2}\varepsilon\int_{t_0}^ta_{1,\mathrm{P}}(f)\,n\,\rmd t,
\]
where we call $n=\left(\mu/a_0^3\right)\!{}^{1/2}$, and the terms of $a_{1,\mathrm{P}}(f)$ with the subindex $j=0$ in (\ref{a1s}) are trivially integrated in the physical time $t$ too because they are free from $f$. Terms with $j\ne0$ in (\ref{a1s}) are integrated in the true anomaly replacing $\rmd t$ by the right side of (\ref{dtf}) with $\tau\equiv{f}$. The integration is easily obtained with the help of modern computer algebra systems, or resorting to the general integration formulas in \cite{Kelly1989}. We obtain
\begin{equation} \label{Il1}
\mathcal{I}= n\left[1+\frac{3}{2}\varepsilon{a}_{1,\mathrm{P}}(f_0)\right](t-t_0)
+\frac{3}{2}\varepsilon\left[\mathcal{I}_\mathrm{P}(f)-\mathcal{I}_\mathrm{P}(f_0)\right],
\end{equation}
in which 
\begin{equation} \label{DIS}
\mathcal{I}_\mathrm{P}=
2\eta(3s^2-2)(f-M+e\sin{f})-s^2\eta\sum_{j=0}^3\tilde{I}_{1,j}\sin(jf+2\omega),
\end{equation}
with the same coefficients $\tilde{I}_{1,j}$ computed before. 
\par

Finally, after replacing (\ref{Il1}) and (\ref{s1}) into (\ref{Moft}), we obtain
\begin{equation} \label{M1}
M_1=M_0+n^*(t-t_0)
+\varepsilon\left[M_\mathrm{P}(f)-M_\mathrm{P}(f_0)\right],
\end{equation}
where
\begin{equation} \label{nstar}
n^*=n\Big[1+\frac{3}{2}\varepsilon{a}_{1,\mathrm{P}}(f_0)-3\varepsilon\eta(3s^2-2)\Big],
\end{equation}
and $M_\mathrm{P}=\beta_{1,\mathrm{P}}(f)+\frac{3}{2}\mathcal{I}_\mathrm{P}(f)$. That is,
\begin{equation}  \label{Ms}
M_\mathrm{P}= \frac{\eta}{4e}(3s^2-2)\sum_{j=1}^3\tilde\omega_{0,j}\sin{j}f
+\frac{\eta}{8e}s^2\sum_{j=-1}^5M^*_j\sin(jf+2\omega),
\end{equation}
with $M^*_{-1}=3e^2$, $M^*_0=2e\left[9-4\eta^2(2+\eta)/(1+\eta)^2\right]$, $M^*_1=3(5e^2+4)$, $M^*_2=0$, $M^*_3=e^2-28$, $M^*_4=-18e$, and $M^*_5=-3e^2$,
which is purely periodic in the mean anomaly. Again, it can be checked that the short-period corrections to the mean anomaly given by $M_\mathrm{P}$ match the correction $dM_s$ in \cite[Eq.~(10)]{Kozai1959} after removing the hidden long-period terms in \cite[Eq.~(12)]{Kozai1959}.
\par

\subsection{Average dynamics}

Because the short-period effects average out in one orbital period, the main dynamical effects of the $J_2$ perturbation are obtained after removing them from (\ref{M1}). We obtain
\begin{equation} \label{M1a}
M_1'= M_0-\varepsilon{M}_\mathrm{P}(f_0)+n^*(t-t_0),
\end{equation}
which shows that, at the precision of the first Picard iteration, the mean anomaly advances, on average, at the rate given by (\ref{nstar}). Remarkably, the average rate described by $n^*$ amends with an additional term the expression commonly accepted in the literature for the average rate of the mean anomaly under $J_2$ perturbations. Namely,
\begin{equation} \label{Mpa}
\frac{\overline{\rmd M}}{\rmd t}=n-nJ_2\frac{R_\Earth^2}{p^2}\frac{3}{4}\eta(3s^2-2),
\end{equation}
cf.~\cite[Eq.~(31)]{Kozai1959}, \cite[Eq.~(11.17)]{Roy2005}, \cite[ Eq.~(3.74)]{Kaula1966}, \cite[Eq.~(5)]{MarkleyJeletic1991}, or \cite[Eq.~(5.25)]{Lara2021}. 
As will be discussed in the examples, the use of $n^*$ instead of $\overline{\rmd M}/\rmd t$ radically improves the analytical propagation of the secular terms by drastically reducing along-track errors.
\par

Proceeding analogously with (\ref{a1})--(\ref{g1}), we obtain
\begin{eqnarray} \label{a1a}
{a}_1' &=& a_0[1-\varepsilon a_{1,\mathrm{P}}(f_0)],
\\ \label{e1a}
e_1' &=& e_0 -\varepsilon e_{1,\mathrm{P}}(f_0),
\\ \label{I1a}
I_1' &=& I_0 -\varepsilon{c}I_{1,\mathrm{P}}(f_0),
\\ \label{h1a} 
\Omega_1' &=& \Omega_0-\varepsilon{c}\Omega_{1,\mathrm{P}}(f_0) -6\varepsilon{c}(M-M_0),
\\ \label{g1a}
\omega_1' &=& \omega_0-\varepsilon\omega_{1,\mathrm{P}}(f_0) -3\varepsilon(5s^2-4)(M-M_0) ,
\end{eqnarray}
that represent the mean $J_2$ dynamics. Now (\ref{a1a})--(\ref{I1a}) show that, in the approximation provided by the first Picard iteration, the semi-major axis, eccentricity, and inclination, remain constant on average. On the other hand, replacing $M$ in (\ref{h1a}) and (\ref{g1a}) by the right side of $M_1'$ in (\ref{M1a}), and differentiating them with respect to the physical time, we obtain
\begin{equation} \label{Oomegaver}
\frac{\rmd \Omega_1'}{\rmd t}=-n^*J_2\frac{R_\Earth^2}{p^2}\frac{3}{2}\cos{I}, \qquad
\frac{\rmd \omega_1'}{\rmd t}=-n^*J_2\frac{R_\Earth^2}{p^2}\frac{3}{4}(5\sin^2I-4).
\end{equation}
where in the $\mathcal{O}(J_2)$ approximation we can replace $n^*$ by $n$. Thus, we obtain the usual result showing that, at the precision of the first Picard iteration, the argument of the perigee and the right ascension of the ascending node undergo secular trends (see \cite[Eqs.~(10.94) and (10.95)]{Battin1999}, for instance). In particular, these average variations show that polar orbits of the $J_2$ problem remain polar except for short-period oscillations, and that inclined orbits of the $J_2$ problem with such inclination that $\sin^2I=4/5$ remain, on average, with fixed line of apsides.
\par

In the particular case in which the time starts at perigee passage $t_0=f_0=M_0=0$, and (\ref{nstar}) turns into
\begin{equation} \label{nstar0}
n^*(0)=n\left[1+\frac{3}{4}J_2\frac{R_\Earth^2}{p^2}\frac{(1+e)^2}{1-e}\left(2-3s^2+3s^2\cos2\omega\right)\right],
\end{equation}
whereas (\ref{a1s})--(\ref{g1s}) and (\ref{Ms}) become
\begin{eqnarray} \label{a10}
a_{1,\mathrm{P}}(0) &=& \frac{2}{\eta^2}\left\{(3s^2-2)\left[\eta^3-(1+e)^3\right]+3(1+e)^3s^2\cos2\omega\right\},
\\ \nonumber
e_{1,\mathrm{P}}(0) &=& \frac{1+e}{1+\eta}\bigg[-(3s^2-2)(2+\eta)(1+e+\eta) \\ \label{e10} &&
+(1+e)\left(e\frac{3+2\eta}{1+\eta}+7\eta+3\right)s^2\cos2\omega\bigg],
\\ \label{G10}
I_{1,\mathrm{P}}(0) &=& 2cs\frac{1+e}{1+\eta}(1+e+2\eta)\cos2\omega,
\\ \label{h10}
\Omega_{1,\mathrm{P}}(0) &=& 2\frac{1+e}{1+\eta}(1+e+2\eta)\sin2\omega,
\\ \nonumber
\omega_{1,\mathrm{P}}(0) &=& \left[15s^2-6-\frac{6s^2-2}{1+\eta}+2\frac{s^2}{e}+(4s^2-2)(2e-\eta)+\frac{s^2}{(1+\eta)^2}\right]\\ \label{g10}
&& \times\sin2\omega,
\\ \label{l10}
M_\mathrm{P}(0) &=&-\eta^3s^2\left[\frac{2+\eta}{(1+\eta)^2}+\frac{2}{e}\right]\sin2\omega,
\end{eqnarray}
in which the symbols in the right side correspond to the initial osculating elements or functions of them.
\par

\subsection{Examples}

\begin{figure}[htb]
\centerline{ \includegraphics[scale=0.72]{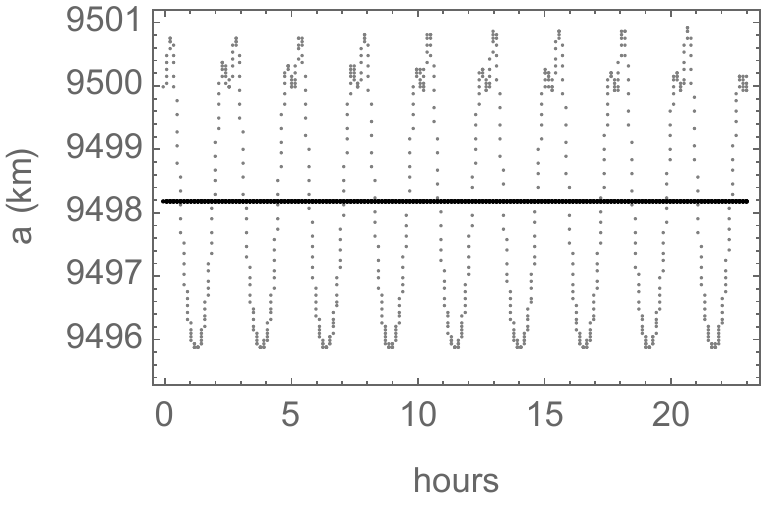} \quad
\includegraphics[scale=0.72]{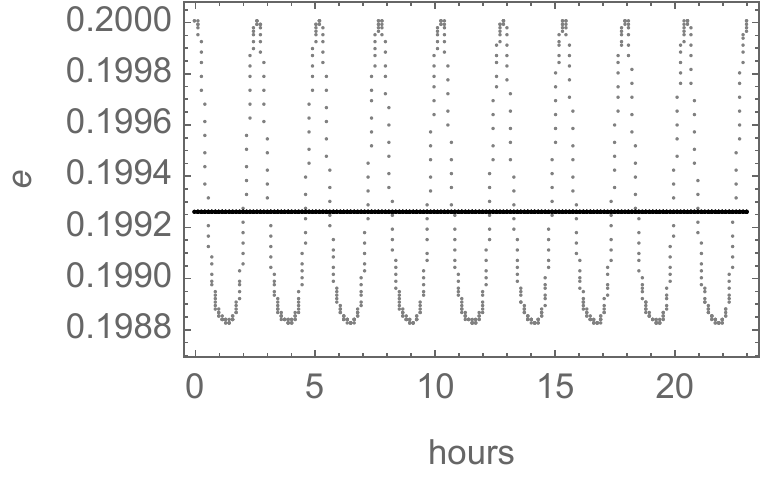} \quad \includegraphics[scale=0.72]{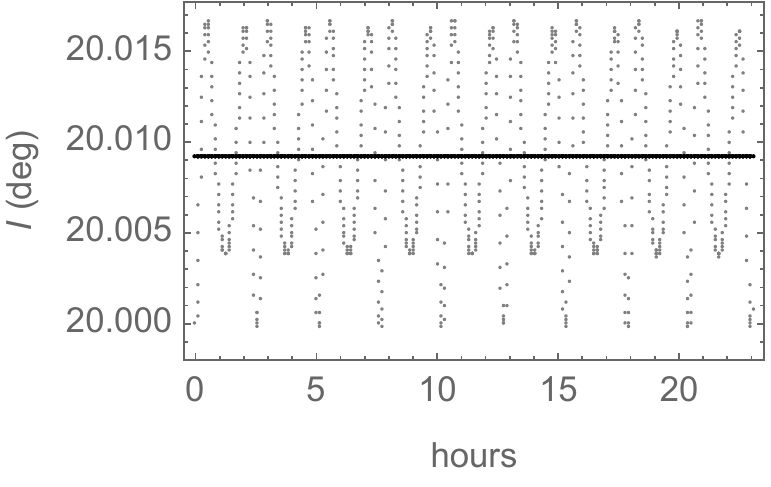} }
\centerline{ \includegraphics[scale=0.72]{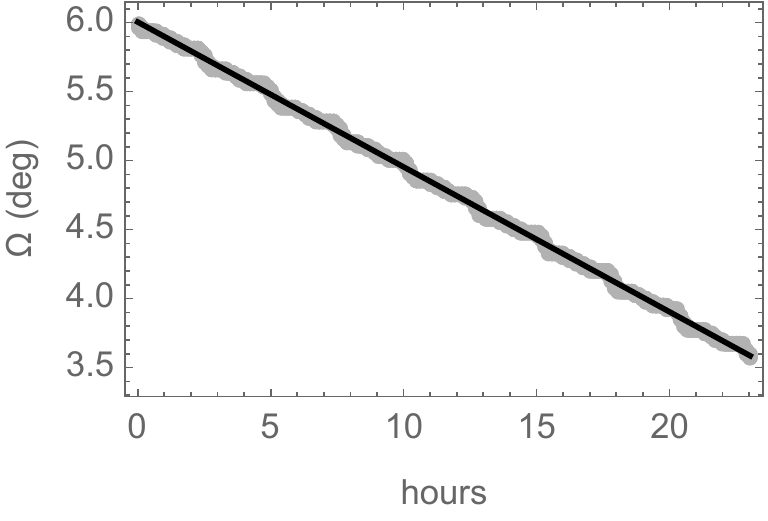} \quad
\includegraphics[scale=0.72]{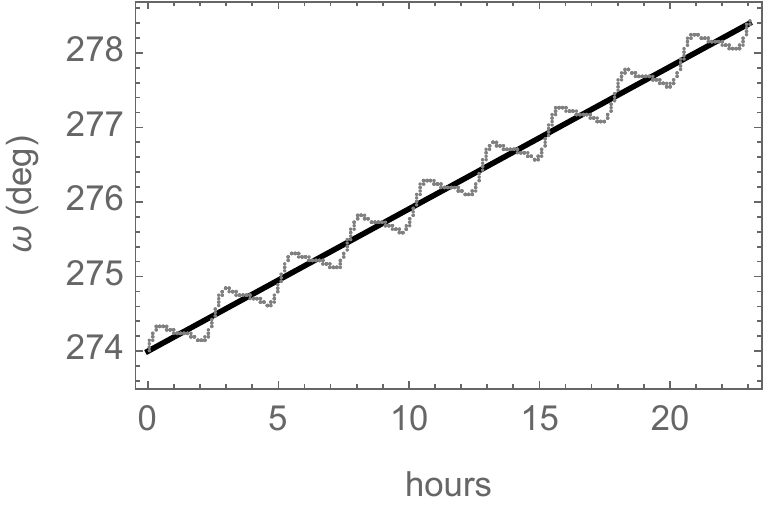} \quad \includegraphics[scale=0.72]{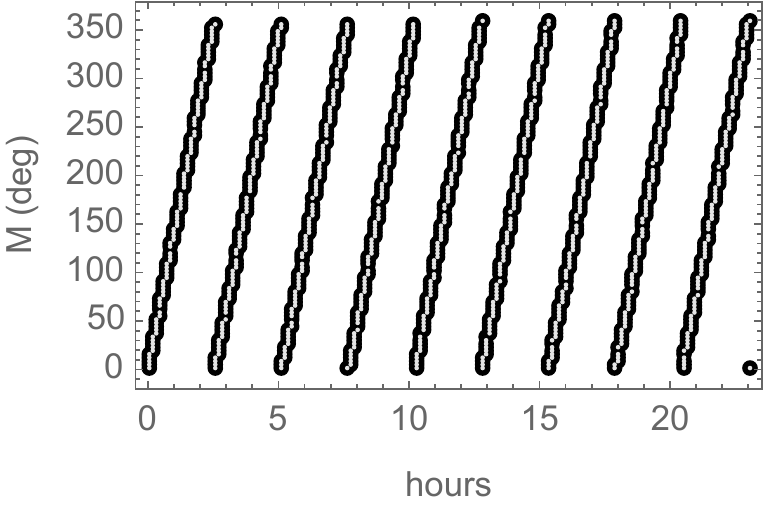} }
\caption{Example elliptic orbit with initial conditions in (\protect\ref{iicc1}). Time history of the mean elements (black) superimposed to the true orbit (gray).
}
\label{f:CA84meantrue}
\end{figure}

To illustrate the performance of the analytical solution obtained by Picard iterations we carry out two examples in each of which the accuracy of the analytical solution is compared with a reference orbit obtained by numerical integration. The first one is the elliptic orbit computed in \cite{CoffeyAlfriend1984}, whose eccentricity guarantees the suitability of the formulation of the analytical solution in orbital elements. Initial conditions of the test orbit are
\begin{equation} \label{iicc1}
a=9500\,\mathrm{km},\quad e=0.2,\quad I=20^\circ,\quad\Omega=6^\circ,\quad\omega=274^\circ,\quad M=0,
\end{equation}
and the propagation is carried out for one day, corresponding to about 9 orbits. The time history of the mean-element dynamics is shown with black lines in Fig.~\ref{f:CA84meantrue}, in which the osculating orbital elements of the true orbit, depicted with gray dots, have been superimposed. As shown in the figure, (\ref{a1a})--(\ref{g1a}) and (\ref{M1a}) effectively capture the average main problem dynamics. More precisely, from (\ref{a1a}) we obtain $a_1'=9498.17$ km, whereas for the average value of the true semi-major axis along the propagation interval $T=24$ hours we obtain
\[
\langle{a}\rangle_T\equiv\frac{1}{T}\int_0^{T}a(t)\,\rmd t=9498.18\,\mathrm{km},
\]
amounting to a relative error of about $10^{-6}$, which is consistent with having neglected $\mathcal{O}(J_2^2)$ effects in the Picard iterations solution. Analogously, from (\ref{e1a}) we obtain $e'_1=0.199256$ whereas $\langle{e}\rangle_T=0.199257$, and the value of $I'_1$ computed from (\ref{I1a}) agrees with $\langle{I}\rangle_T$ to six significant digits when given in radians.
\par

The accuracy of the mean frequencies provided by the analytical solution are estimated by comparing the values given by (\ref{Oomegaver}) and (\ref{M1a}) with respect to corresponding linear fits to the time histories of $\Omega$, $\omega$, and $M$, obtained from the numerical integration. While this kind of comparison may be a little tricky because long-period terms existing in the true orbit may be corrupting the fit for this short propagation interval, we find the results illustrative anyway. We obtained that the slope of the linear fit to $M$ agrees with its mean rate given by (\ref{nstar}) to six significant digits. However, we only find agreement between the predicted mean variations of $\Omega$ and $\omega$ in (\ref{Oomegaver}) and the variation rates of the corresponding linear fits to their true evolution in the three and two first significant digits, respectively. Still, since these mean frequencies are already $\mathcal{O}(J_2)$, the errors between the true average and the predicted mean values remain reasonably small. In particular, for the mean rate of $\omega$ we found an error of about $19$ arc seconds per orbit. 
\par

\begin{figure}[htb]
\centerline{ \includegraphics[scale=0.71]{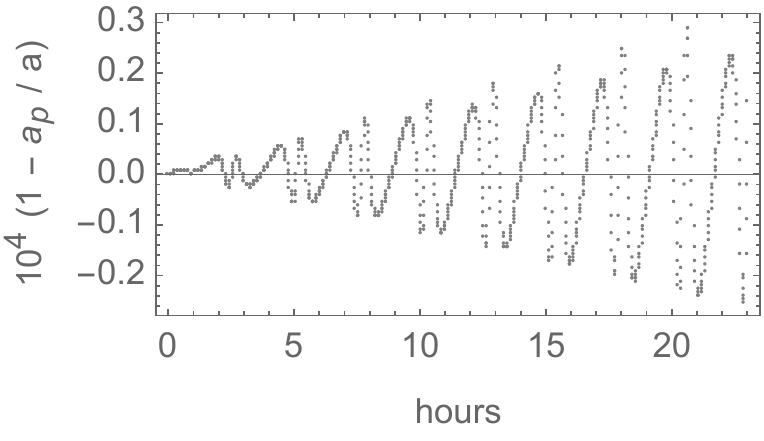} \quad
\includegraphics[scale=0.72]{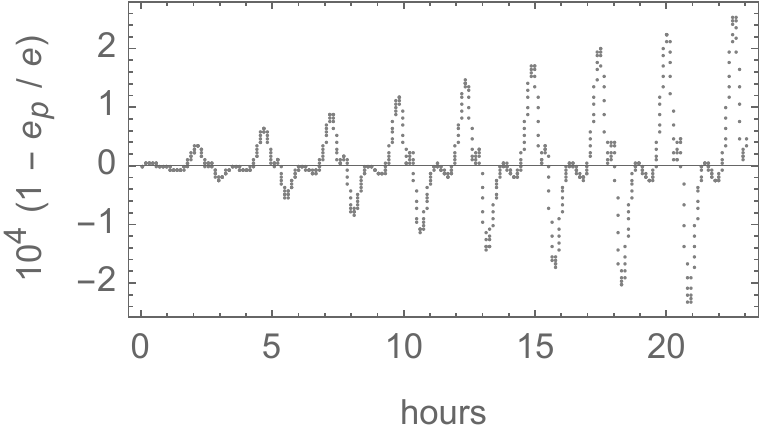} \quad \includegraphics[scale=0.72]{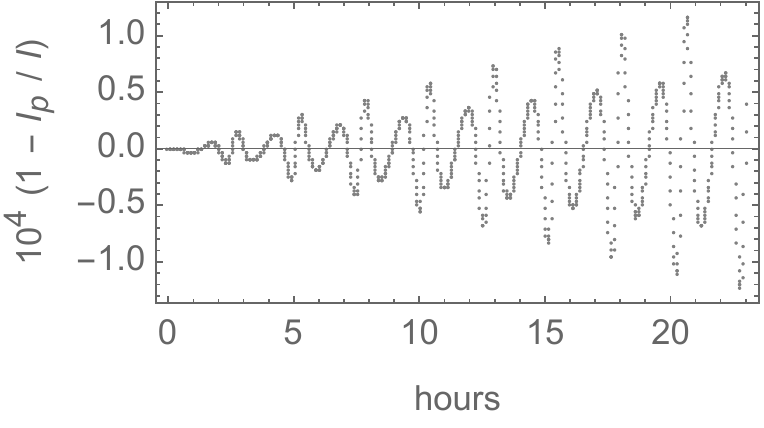} }
\centerline{ \includegraphics[scale=0.72]{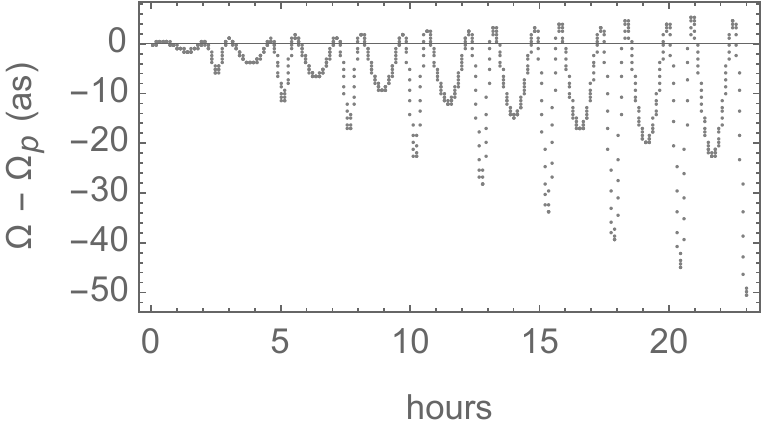} \quad
\includegraphics[scale=0.72]{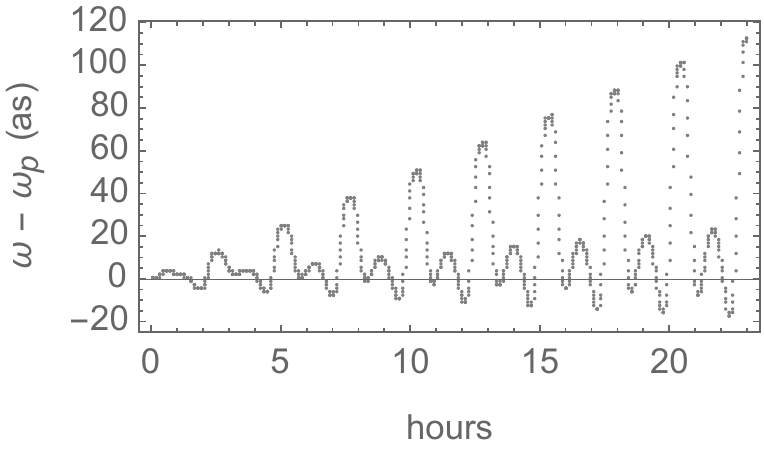} \quad \includegraphics[scale=0.72]{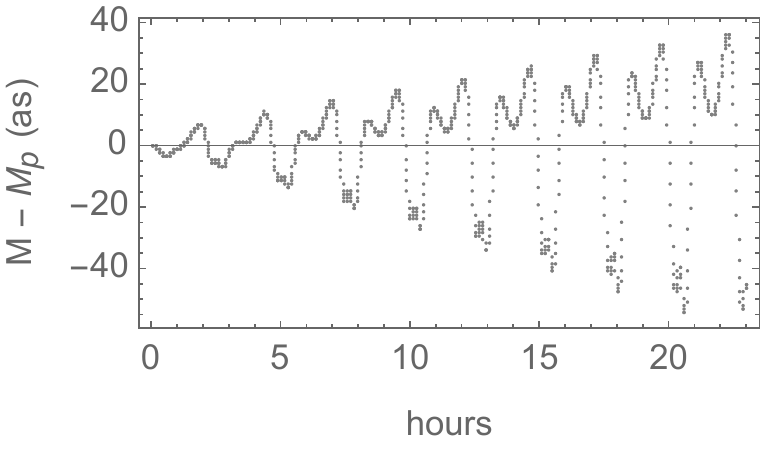} \quad }
\caption{Time history of the errors of the first Picard iteration of the elliptic orbit in (\protect\ref{iicc1})}
\label{f:CA84errors1}
\end{figure}

The errors of the osculating Keplerian elements obtained with the analytical solution along the one-day propagation interval are shown in Fig.~\ref{f:CA84errors1}, where the subindex $p$ denotes the Picard iterations solution. The displayed errors are relative values for $a$, $e$, and $I$, and absolute errors in arc seconds for $\Omega$, $\omega$ and $M$. Note the linear growing of the amplitude of the short-periodic errors in all cases, from the residual small amplitude at the beginning of the interval, due to the restriction of the solution to the first order of $J_2$, to about one order of magnitude larger at the end of the first day. This is mostly due to the inability of the first Picard iteration to model long-period effects, whose lower frequency thus modulates the short-period errors. Besides, as shown in the plots of the bottom row of Fig.~\ref{f:CA84errors1}, the errors of the rotating angles remain of comparable magnitude. The mild behavior of the mean anomaly is due to the additional term in (\ref{nstar}) for the secular part of mean motion $n^*$ when compared with the usual secular rate in (\ref{M1a}). We checked that if the later is used instead of the former, the error in the mean anomaly grows by about two orders of magnitude at the end of the one-day interval in the current example, reaching an amplitude close to one degree, which translates into about two hundred km along-track as opposed to the km level obtained when using $n^*$.
\par

For a second example we take a low-eccentricity orbit with the initial conditions
\begin{equation} \label{iicc2}
a=7707.27\,\mathrm{km},\; e=0.01,\; I=63.4^\circ,\;\Omega=180^\circ,\;\omega=270^\circ,\; M=0,
\end{equation}
corresponding to the configuration of the popular Topex orbit \cite{Frauenholzetal1998}, and the propagation is likewise carried out up to one day, which now amounts to about 13 orbits. We checked that the agreement between the mean elements dynamics and the average dynamics of the true orbit preserves in each case an analogous number of digits to the previous example. Therefore, further details are not given in this regard, save for repeating the comment that along-track error of hundreds of km obtained when using the customary secular ratio in (\ref{Mpa}) are again reduced to the km level when using the mean motion $n^*$ in (\ref{nstar}).
\par

\begin{figure}[htb]
\centerline{ \includegraphics[scale=0.71]{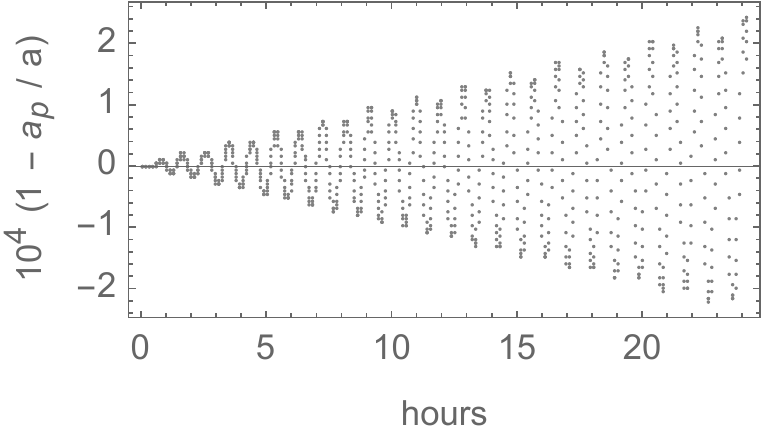} \quad
\includegraphics[scale=0.72]{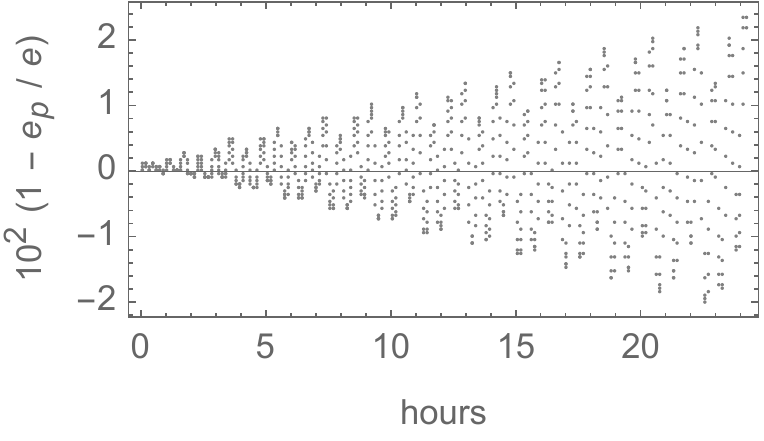} \quad \includegraphics[scale=0.72]{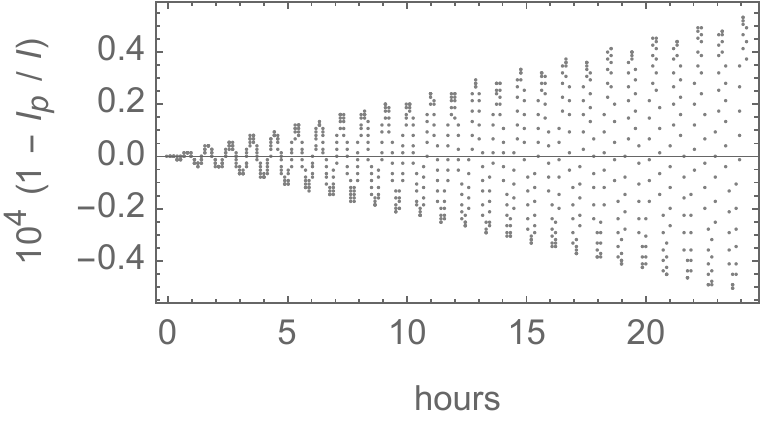} }
\centerline{ \includegraphics[scale=0.72]{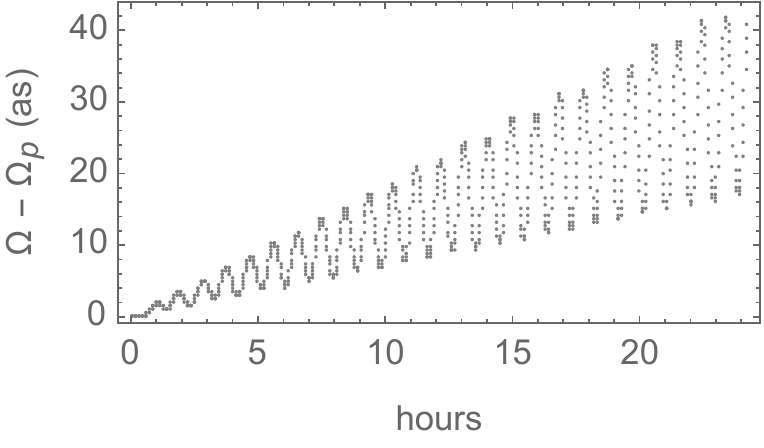} \quad \includegraphics[scale=0.72]{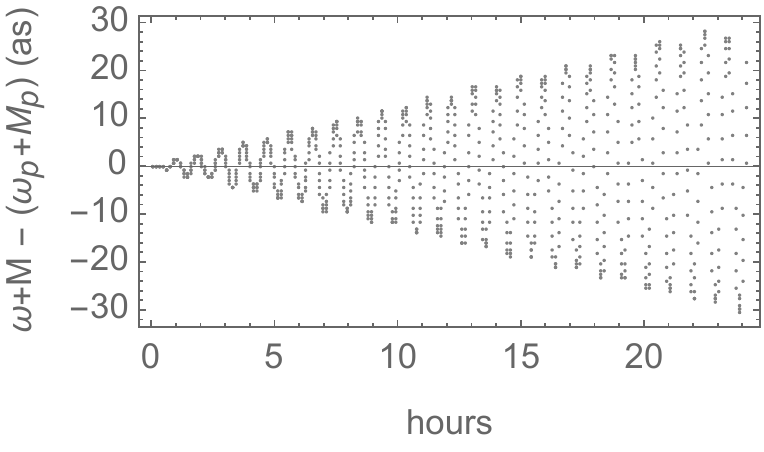} }
\caption{Topex orbit. Time history of the errors of the first Picard iteration.
}
\label{f:Topexerrors}
\end{figure}

Because the eccentricity is now small, it clearly enhances the  effect of the $J_2$ perturbation on the argument of the periapsis and the mean anomaly due to the appearance of factors $J_2/e$ in (\ref{g1s}) and (\ref{Ms}), respectively. This is exactly the case, and we found that while in the previous example the errors of the argument of the perigee and the mean anomaly had amplitudes of tens of arc seconds (plots in the last row of Fig.~\ref{f:CA84errors1}), they have now amplitudes in the degree level for the Topex orbit. However, this poor modeling of low-eccentricity orbits is just a consequence of our choice of the Keplerian elements for the formulation of the analytical solution. As expected from (\ref{dFdt}), which is free from offending divisors, the trouble is easily amended by replacing $\omega$ and $M$ by the mean distance to the node $F=\omega+M$. This is illustrated in Fig.~(\ref{f:Topexerrors}), where we show the time history of the errors of the first Picard iteration with respect to the true, numerically integrated solution. The plots in the two first rows of Fig.~(\ref{f:Topexerrors}) show the similar behavior of the relative errors of $a$, $e$, $I$, and $\Omega$ to those of the previous example. In the bottom plot of the figure we replaced the errors in the argument of the perigee and the mean anomaly by their combination, in this way bringing back the errors of the analytical solution to the arc second level, like in the previous example.
\par

\section{Second Picard iteration}

The first iteration of Picard's method misses long-period perturbations that clearly show up in propagation intervals of just a few orbits. A refinement of the analytical solution that captures non-resonant long-period effects of the dynamics is obtained, up to the first order of $J_2$, by means of an additional iteration of (\ref{newrap}).
\par

To this effect, the orbital elements given by (\ref{a1})--(\ref{s1}) are replaced in the right side of the variation equations (\ref{dxdf}), which then only depend on the fictitious time $f$ and the initial orbital elements, and hence are again solved by indefinite integration. To carry out the integration analytically in closed form of the eccentricity, we expand the right side of the equations in powers of $J_2$, which we truncate to the first order. However, to avoid the appearance of spurious mixed secular-periodic terms, we do not expand trigonometric terms in what respects to the argument $\omega-3\varepsilon(5s^2-4)f$, which results from the combination of the secular term in (\ref{g1}) with the one involving the equation of the center in (\ref{g1s}).
\par

To the first order of $J_2$, the whole procedure is equivalent to replacing $M$ by $n^*t$ and $\omega$ by $\omega(f)\equiv\omega_0-3\varepsilon(5s^2-4){f}$ in the solution (\ref{a1})--(\ref{s1}) given by the first Picard iteration. In this way, the second iteration of Picard's method yields the new solution
\begin{eqnarray} \label{a2}
a_2=& \; a_0 +a_0\varepsilon\left[a_{1,\mathrm{P}}(f,\omega(f))-a_{1,\mathrm{P}}(f_0,\omega_0)\right],
\\ \label{e2}
e_2=& \; e_0 +\varepsilon\left[e_{1,\mathrm{P}}(f,\omega(f))-e_{1,\mathrm{P}}(f_0,\omega_0)\right],
\\ \label{I2}
I_2=& \; I_0 +\varepsilon{c}\left[I_{1,\mathrm{P}}(f,\omega(f))-I_{1,\mathrm{P}}(f_0,\omega_0)\right],
\\ \label{h2} 
\Omega_2=& \; \Omega_0 -6\varepsilon{c}n^*(t-t_0) +\varepsilon{c}\left[\Omega_{1,\mathrm{P}}(f,\omega(f))-\Omega_{1,\mathrm{P}}(f_0,\omega_0)\right],
\\ \label{g2}
\omega_2=& \; \omega_0 -3\varepsilon(5s^2-4)n^*(t-t_0) +\varepsilon\left[\omega_{1,\mathrm{P}}(f,\omega(f))-\omega_{1,\mathrm{P}}(f_0,\omega_0)\right],
\\ \label{s2}
\beta_2=& \; -3\varepsilon\eta(3s^2-2)n^*(t-t_0) +\varepsilon\left[\beta_{1,\mathrm{P}}(f,\omega(f))-\beta_{1,\mathrm{P}}(f_0,\omega_0)\right],
\end{eqnarray}
where the modifications introduced by the second iteration in the frequencies of the arguments involving $\omega$ yield now long-period terms in addition to the short-period ones. 
\par

The solution (\ref{M1}) of mean anomaly should be analogously improved when solving (\ref{Moft}) with $k=2$. However, in this process we find integrals whose closed form solution is not known. Therefore, we keep for the integrand the value (\ref{Il1}) obtained in the first iteration, yet the resulting value of $M$ is modified by replacing, as before, $M=n^*t$ and $\omega=\omega(f)$ in (\ref{DIS}), and the use of the refined solution for $\beta$ in (\ref{s2}). On the other hand, the value of the fictitious time $f=f(t)$ needed in the evaluation of the refined solution is again obtained from (\ref{foft}). Like in the previous iteration the solution $f$ is obtained up to $\mathcal{O}(J_2)$ by solving Kepler's equation in which, instead of using the mean motion corresponding to the initial osculating semimajor axis, we replace $n=\left(\mu/a_1^3\right)\!{}^{1/2}$. In practice, it is enough to replace this value by the secular variation $n^*$ given in (\ref{nstar}).
\par

\begin{figure}[htb]
\centerline{ \includegraphics[scale=0.72]{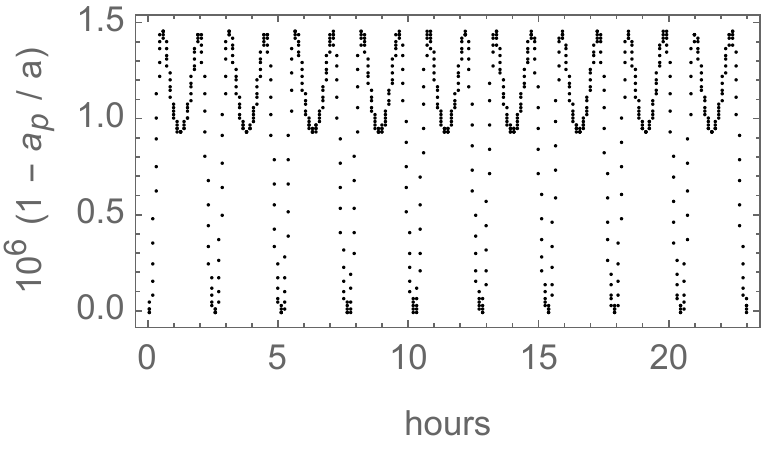} \quad
\includegraphics[scale=0.71]{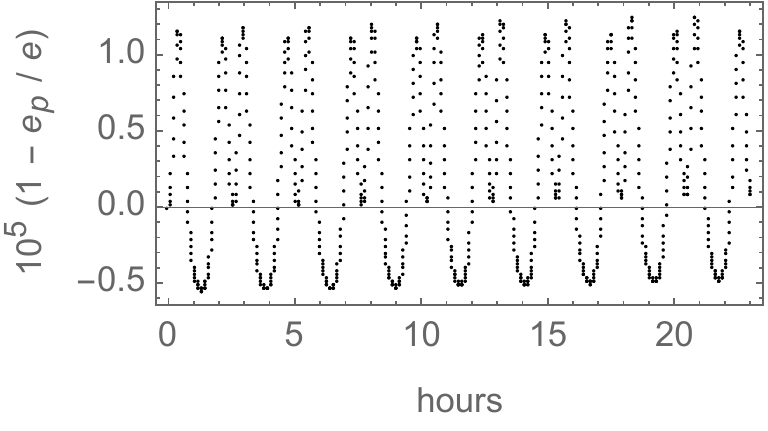} \quad \includegraphics[scale=0.72]{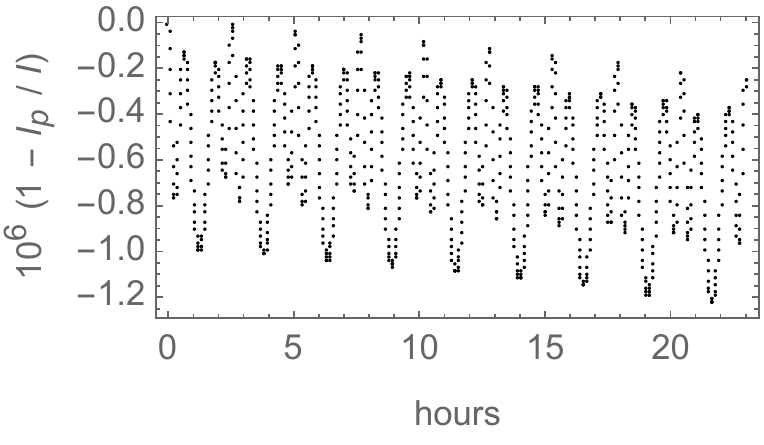} }
\centerline{ \includegraphics[scale=0.72]{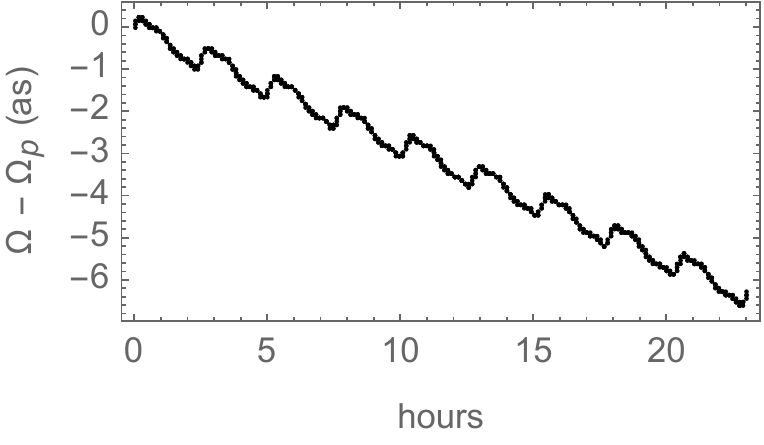} \quad
\includegraphics[scale=0.72]{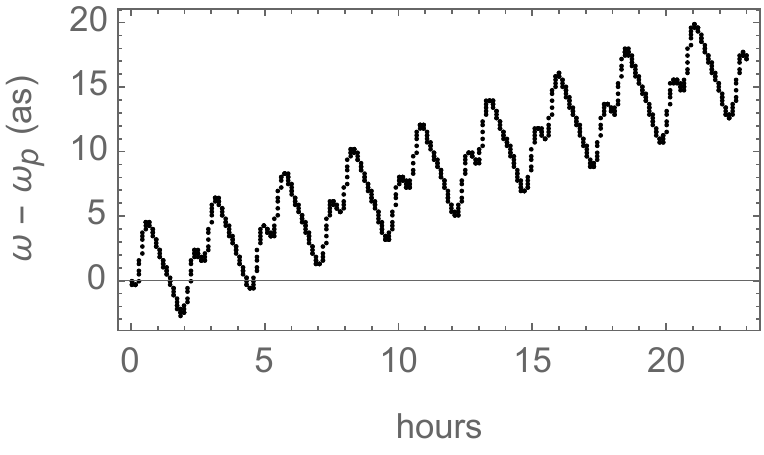} \quad \includegraphics[scale=0.72]{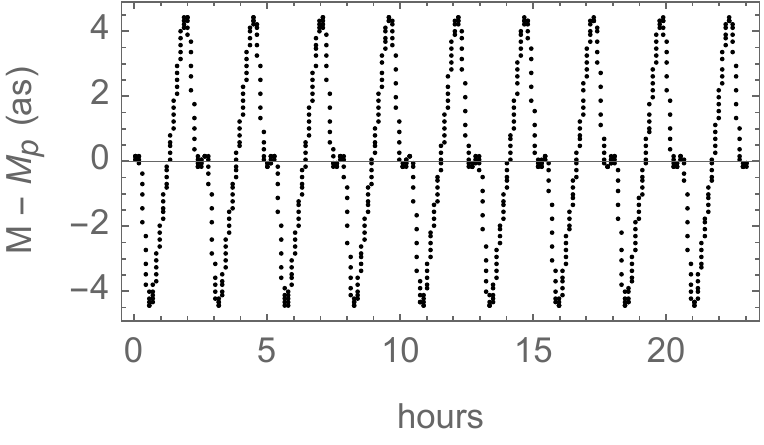} }
\caption{Time history of the errors of the second Picard iteration with respect to the true elliptic orbit in (\protect\ref{iicc1}).
}
\label{f:CA84errors2}
\end{figure}

\begin{figure}[htb]
\centerline{ \includegraphics[scale=0.72]{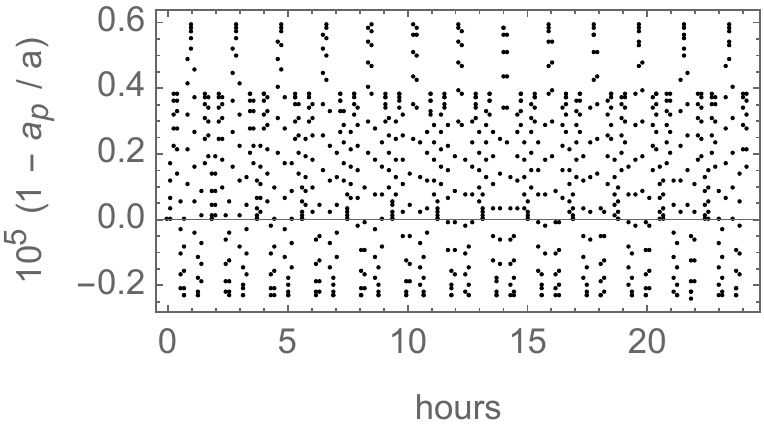} \quad
\includegraphics[scale=0.71]{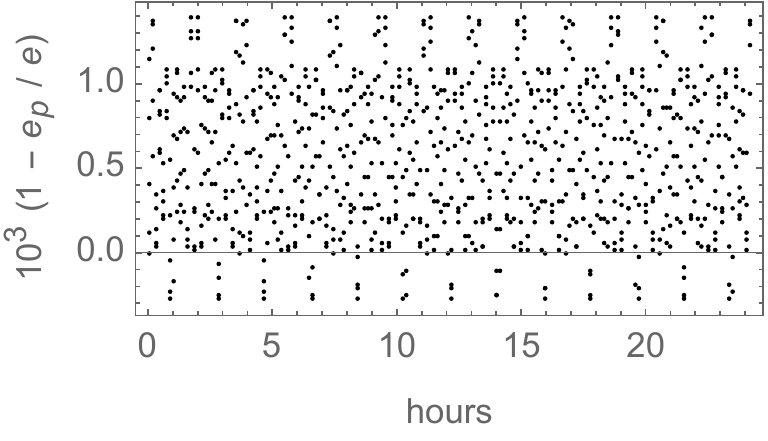} \quad \includegraphics[scale=0.72]{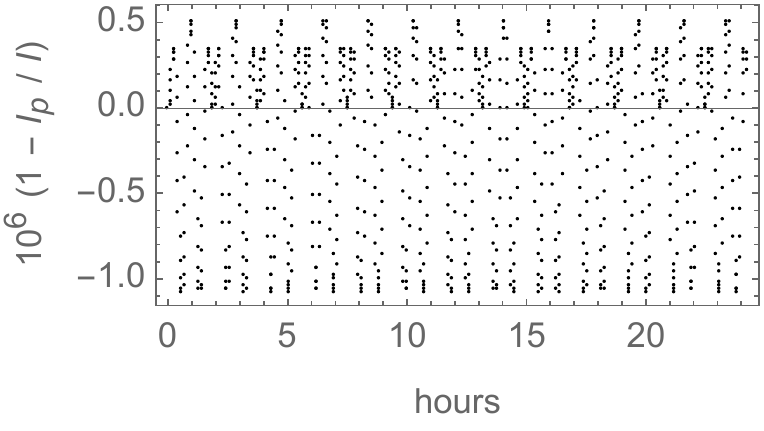} }
\centerline{
\includegraphics[scale=0.72]{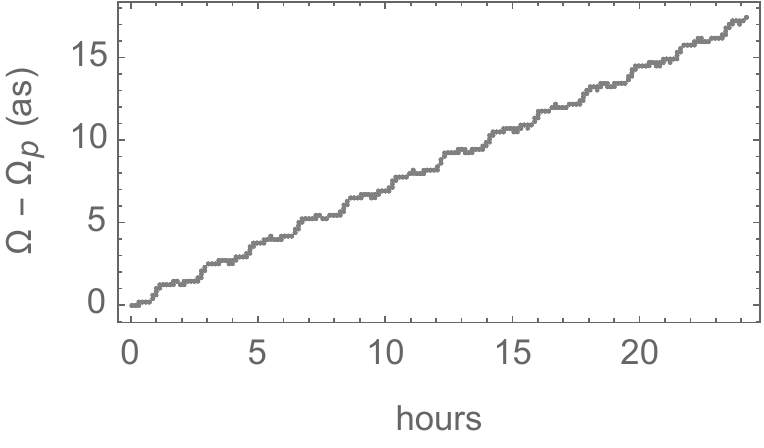} \quad \includegraphics[scale=0.72]{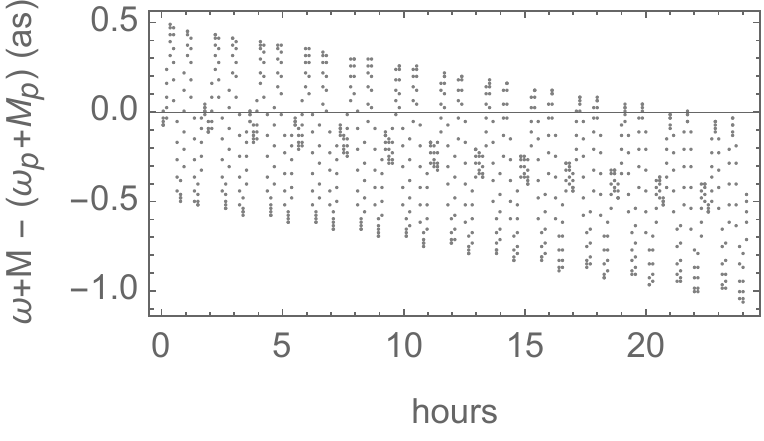} }
\caption{Time history of the errors of the second Picard iteration with respect to the true Topex orbit
}
\label{f:Topexerrors2}
\end{figure}

The errors of the Keplerian elements obtained with the second Picard iteration along the one-day propagation interval are shown in Fig.~\ref{f:CA84errors2} for the eccentric orbit case, and in Fig.~\ref{f:Topexerrors2} for the Topex orbit. While the errors start with the same amplitudes as those of the first Picard iteration, the influence of the long-period terms, which are not captured with the first Picard iteration, become now evident. Now, the amplitudes of the errors remain mostly constant along the propagation, thus improving the errors with respect to the first Picard iteration by about one order of magnitude at the end of the one-day propagation interval. Still, remaining secular components,  which are a consequence of the truncation of the solution to first order effects of $J_2$, are clearly observed in the last rows of Figs.~\ref{f:CA84errors2} and Fig.~\ref{f:Topexerrors2}, and, most notably, in the time history of the errors of $\Omega$, $\omega$ and $M$, or $F$, thus showing the limitations of the Picard iterations solution in the modeling of the $J_2$ dynamics.
\par

%

\section{Conclusions}

A simple analytical solution of the main problem of artificial satellite theory has been computed in closed form based on the classical method of Picard iterations that guarantees the existence and unicity of the solution in some range of the independent variable. The appearance of spurious mixed secular-periodic term in the repeated iterations of Picard's method has been avoided proceeding in the style of the popular linear theory of Kaula. In this way, the solution, which depends only on elementary functions, can seize non-resonant, long-period terms of the main problem dynamics in addition to the usual short-period terms.
\par

The new solution is accurate only to $\mathcal{O}(J_2)$ effects, and hence limited in application. Notwithstanding, this is the same accuracy provided by the majority of intermediary orbits of the main problem, over which the new solution may be preferred for its simplicity and direct insight. In particular, it discloses a secular linear trend in the mean anomaly that is different from the standard result in the literature. This improved value of the secular component of the mean motion is a consequence of approaching the integration of the variation of parameters equations in a fictitious time, rather than the physical one, whose rate is proportional to the time variation of the true anomaly. 
\par

%

\ack
Partial support by the Spanish State Research Agency and the European Regional Development Fund (Project PID2020-112576GB-C22, AEI/ERDF, EU) is recognized. 

\end{document}